\documentclass[final,conference]{IEEEtran}
\IEEEoverridecommandlockouts
\usepackage{cite}
\usepackage{amsmath,amssymb,amsfonts}
\usepackage{algorithmic}
\usepackage{graphicx}
\usepackage{textcomp}
\usepackage{xcolor}
\usepackage{colortbl}
\usepackage{hyperref}
\usepackage{todo}
\usepackage{tikz}
\usepackage{balance}

\usetikzlibrary{arrows}
\usetikzlibrary{shapes}

\definecolor{blue}{RGB}{30,63,130}
\definecolor{white}{RGB}{255,255,255}

\newcommand*\circled[2]{\texttt{\tikz[baseline=(char.base)]{
        \node[fill=#2,shape=circle,draw,minimum size=3.6mm, inner sep=0pt] (char)
        {\textcolor{white}{#1}};}}}

\def\BibTeX{{\rm B\kern-.05em{\sc i\kern-.025em b}\kern-.08em
    T\kern-.1667em\lower.7ex\hbox{E}\kern-.125emX}}
\begin{document}

\title{MANAi - An IntelliJ Plugin for Software Energy Consumption Profiling}

\author{\IEEEauthorblockN{Andreas Schuler\IEEEauthorrefmark{1}\IEEEauthorrefmark{2} and Gabriele Kotsis\IEEEauthorrefmark{1}}
\IEEEauthorblockA{\IEEEauthorrefmark{1} Department of Telecooperation
\textit{Johannes Kepler University} Linz, Austria}
\IEEEauthorblockA{ \IEEEauthorrefmark{2}Advanced Information Systems and Technology (AIST)
\textit{University of Applied Sciences Upper Austria} Hagenberg, Austria}
andreas.schuler@fh-hagenberg.at, gabriele.kotsis@jku.at
}

\maketitle

\begin{abstract}
Developing energy-efficient software solutions 
is a tedious task. We need both, the awareness that 
energy-efficiency plays a key role in modern software
development and the tools and techniques to support 
stakeholders involved in the software development
lifecycle. So, we present the MANAi plugin
which helps to make energy consumption of unit test methods explicit 
by providing visual feedback as a plugin to the \emph{Integrated 
Development Environment (IDE)} IntelliJ. Our tool is intended to 
bring software energy consumption into the limelight
as an important non-functional quality aspect in software development. 
Furthermore, with MANAi we provide a tool that eases the  
process of software energy experiments for a broad range of users 
from academia to industry.
\end{abstract}


\section{Introduction}
Meeting sustainable design goals when developing software can play a crucial role in shaping our future. 
Given the ever-increasing demand for digitalization, software will continue to support or even take over 
different areas of our daily life. As a result, developing software faces ongoing challenges in 
providing sustainable and in particular energy-efficient software solutions. 
In supporting researchers and practitioners alike to tackle these challenges, software 
engineering research has provided considerable contributions to comprehend how 
energy is dissipated in software and how software design affects energy consumption. 
These contributions range from the implications of software design choices, to the 
development of energy optimized applications \cite{141,518,142,147,011,954,009,Oliveira2021a,912,Pereira2021,9463126,10.1145/3382494.3410688}.
With our work, MANAi, we complement the body of existing methods with a 
tool that facilitates access to software energy profiling for a broad 
audience, ranging from computer science students and researchers to practitioners. 

\section{MANAi}
\begin{figure}[!h]
    \vspace{-1.0em}
    \centering
    \includegraphics[width=0.47\textwidth]{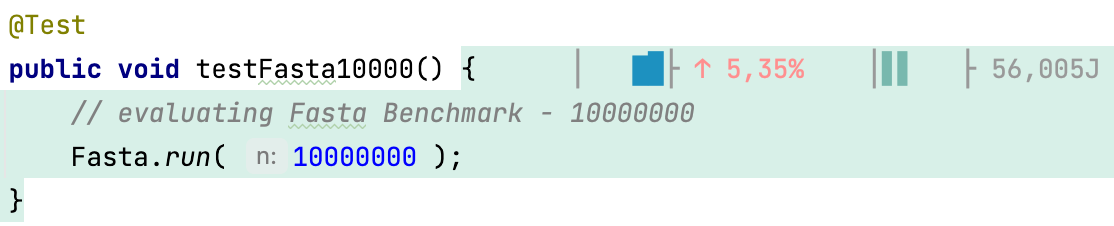}
    \caption{Depiction of a unit test method after execution via MANAi. 
    The method is color coded and inline diagrams provide feedback on its energy characteristics.}
    \label{fig:example}
    \vspace{-1.5em}
\end{figure}

The main goal of MANAi is to shed light on the energy implications of software design 
choices during development right where it is needed in the course of an IDE. 
MANAi shares some design considerations with the 
work from Liu et al. \cite{Liu2015DataOrientedCO}, 
mainly leveraging \emph{Intel's Running Average Power Limit (RAPL)} at its core to 
obtain energy readings. However, we take the idea of \emph{jRAPL} \cite{Liu2015DataOrientedCO} one 
step further and combine data obtained via Intel's RAPL interface with the possibilities of 
expressive visualizations in IDEA's IntelliJ IDE. 
Other existing plugins like EcoAndroid by Cuoto et al. \cite{525} or \emph{Automated Android Energy-Efficiency InspectiON (AEON)} 
maintained by the Software Engineering Lab at George Mason University, targeted for the Android platform, 
share some design considerations regarding visual feedback right within an IDE. 
To give the reader an impression, of what MANAi can do, refer to the illustrative example 
in \figurename{\ref{fig:example}}. In particular, \figurename{\ref{fig:example}} outlines a unit 
test method after MANAi has obtained energy readings during its execution. 
The generated inline information 
attributed to the method provides users with insights, 
not only on the current energy feedback of a method, but furthermore it
allows a user to get insights how the energy characteristics of the examined 
method has evolved, considering it has undergone 
several changes.

\subsection{Architecture}
MANAi is composed by two core modules, 
the \emph{Infrastructure (IFS)} and the \emph{Experiment Environment (ExEn)} module. 
The \emph{IFS} module leverages the IntelliJ 
Platform SDK\footnote{Intellij Platform SDK -- https://tinyurl.com/2p9ed8za} 
and provides components for energy experiment 
definition, data acquisition, visualization and configuration. 
The \emph{ExEn} module facilitates 
the maven-plugin mechanism\footnote{Maven Plugin Development -- https://tinyurl.com/4v637zdu} 
and is responsible to execute energy experiments defined and configured 
via the \emph{IFS} module and report the recorded data obtained using Intel's RAPL. 
Through the \emph{IFS} module, a user can configure and define energy 
experiments within the development environment. The module 
further instructs the \emph{ExEn} 
module to execute defined energy experiments and 
report recorded energy samples back to the user, e.g. 
through visual inline feedback attributed to the executed 
method (cf. \figurename{\ref{fig:example}}). If required, the \emph{ExEN}
module can be used in a headless manner (i.e. without IntelliJ) 
to allow scenarios which are independent of the IDE, e.g. as a step 
in a continuous integration pipeline. Additionally, the \emph{ExEN} module
has been designed with extensibility in mind, to ease future integration
of additional methods to obtain energy readings next to RAPL.

\begin{figure}
    \centering
    \includegraphics[width=0.47\textwidth]{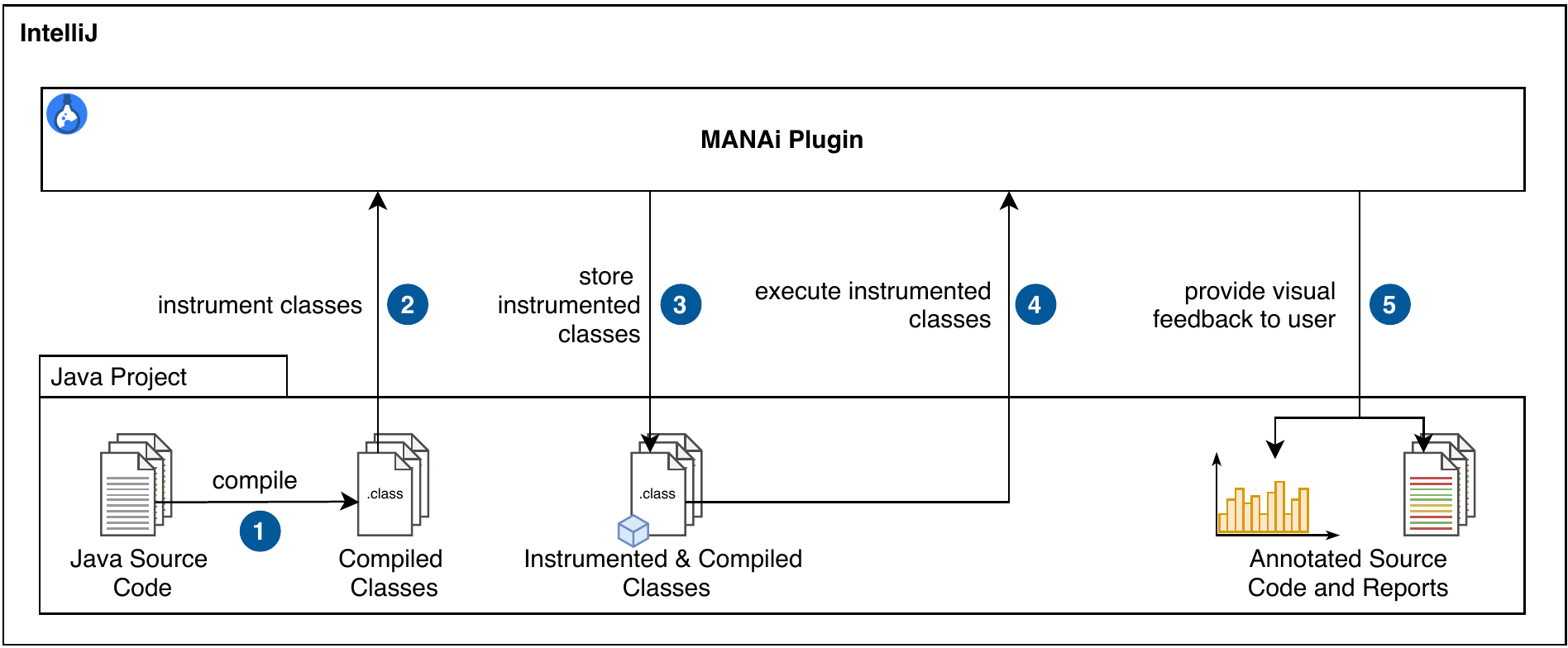}
    \caption{Workflow from Java source code to per method attributed energy consumption data.}
    \label{fig:workflow}

\vspace{-2em}
\end{figure}

The general workflow behind MANAi is described in \figurename{\ref{fig:workflow}}. MANAi
hooks into the compile process of Java source code and
locates unit test classes as possible candidates for obtaining 
energy consumption readings \circled{1}{blue}. Entry points for an energy experiment 
in MANAi are unit test methods. MANAi automatically
detects these unit test methods inside 
a Java project in IntelliJ and tags them as potential 
energy experiment candidates. These candidates are instrumented \circled{2}{blue} to obtain 
energy readings during their execution. 
The instrumented classes are stored \circled{3}{blue} within the currently 
opened project. Using IntelliJ's internal program runner infrastructure, 
the instrumented classes and unit test methods are executed \circled{4}{blue}, 
and the energy data is being recorded. Finally, \circled{5}{blue} the recorded data 
is being visualized within the development environment. As part 
of our replication package \cite{schuler_andreas_icts}, we provide a short video that 
shows how this workflow presents itself from a user perspective. 

\subsection{Features}
One of the key features of MANAi is that it offers different views and
windows that provide expressive visual feedback to users about the energy characteristics
of their Java projects. In what follows, we give a brief overview of these particular features:

\begin{figure}[!b]
    \vspace{-2.0em}
    \centering
    \includegraphics[width=0.47\textwidth]{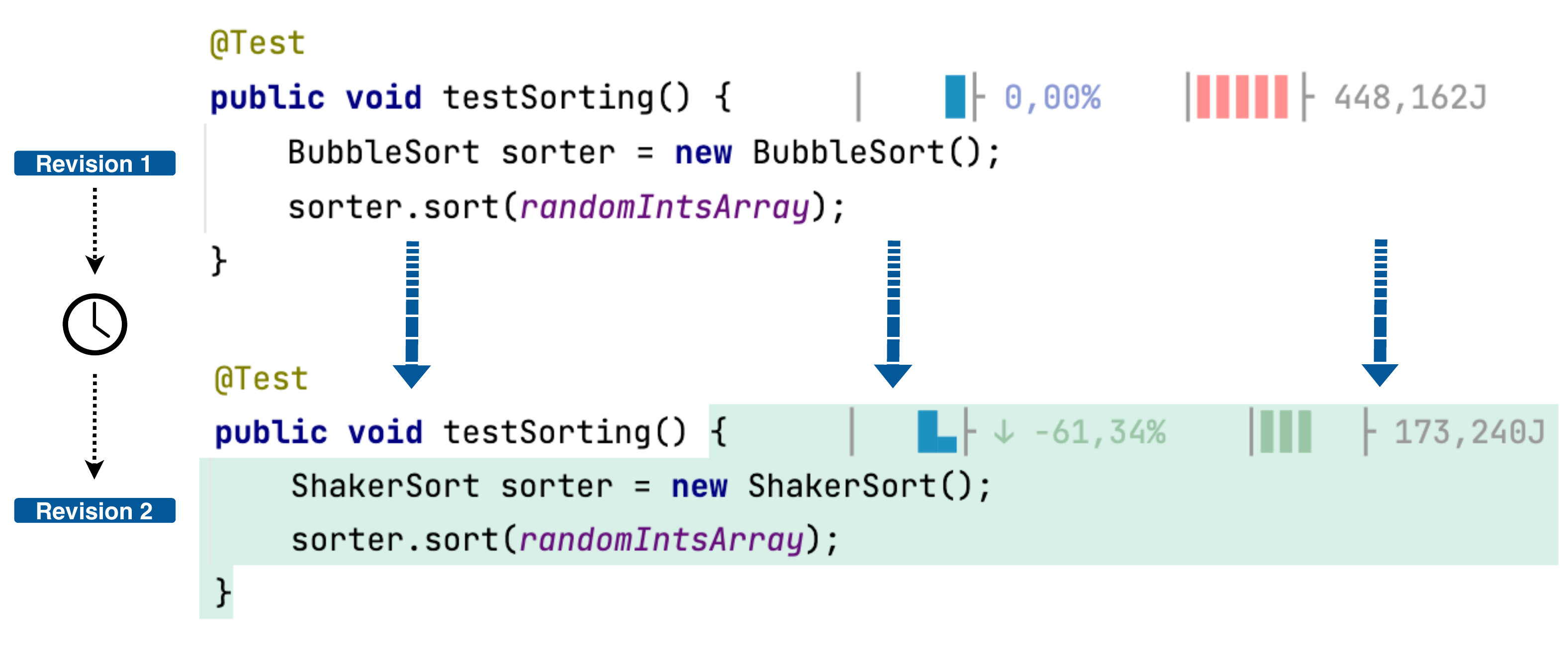}
    \caption{The \emph{Line Annotator} provides visual feedback on the energy characteristics of a method and its evolution amongst different revisions.}
    \label{fig:code}
    \vspace{-2.0em}
\end{figure}

\paragraph{Experiment Runner} The \emph{Experiment Runner} facilitates IntelliJ's internal 
program runner infrastructure to define and execute energy experiments from within the IDE. Defining 
an experiment consists of setting the sampling rate, the number of samples being collected and 
selecting the unit test classes and methods being executed. This allows for replication of experiments, 
as once an energy experiment is defined, it can be rerun multiple times. 
\paragraph{Line Annotator} The \emph{Line Annotator} allows for a fast an easy to comprehend 
visual feedback of the energy characteristics of a unit test method directly within the source code. 
Next to the last energy consumption recorded for a particular unit test, the inline charts highlight 
how the method has evolved, considering their energy attributes. This further allows comparing 
the energy characteristics of different unit test methods. Furthermore, users are able to 
inspect how a recent change to a method affects its energy consumption
compared to recorded energy data of previous revisions. 
For example, let us assume we have a method that applies a rather naive approach to sorting an array of 
integer values, which we call \emph{Revision 1} as depicted in Fig. \ref{fig:code}. 
Now, if we adapt the implementation of \emph{Revision 1} with a slightly more
efficient sorting algorithm in \emph{Revision 2}, the \emph{Line Annotator} highlights 
if the applied changes led to an in- or decrease in energy consumption (cf. Fig. \ref{fig:code}).

\begin{figure}[!h]
    \vspace{-1em}
    \centering
    \includegraphics[width=0.46\textwidth]{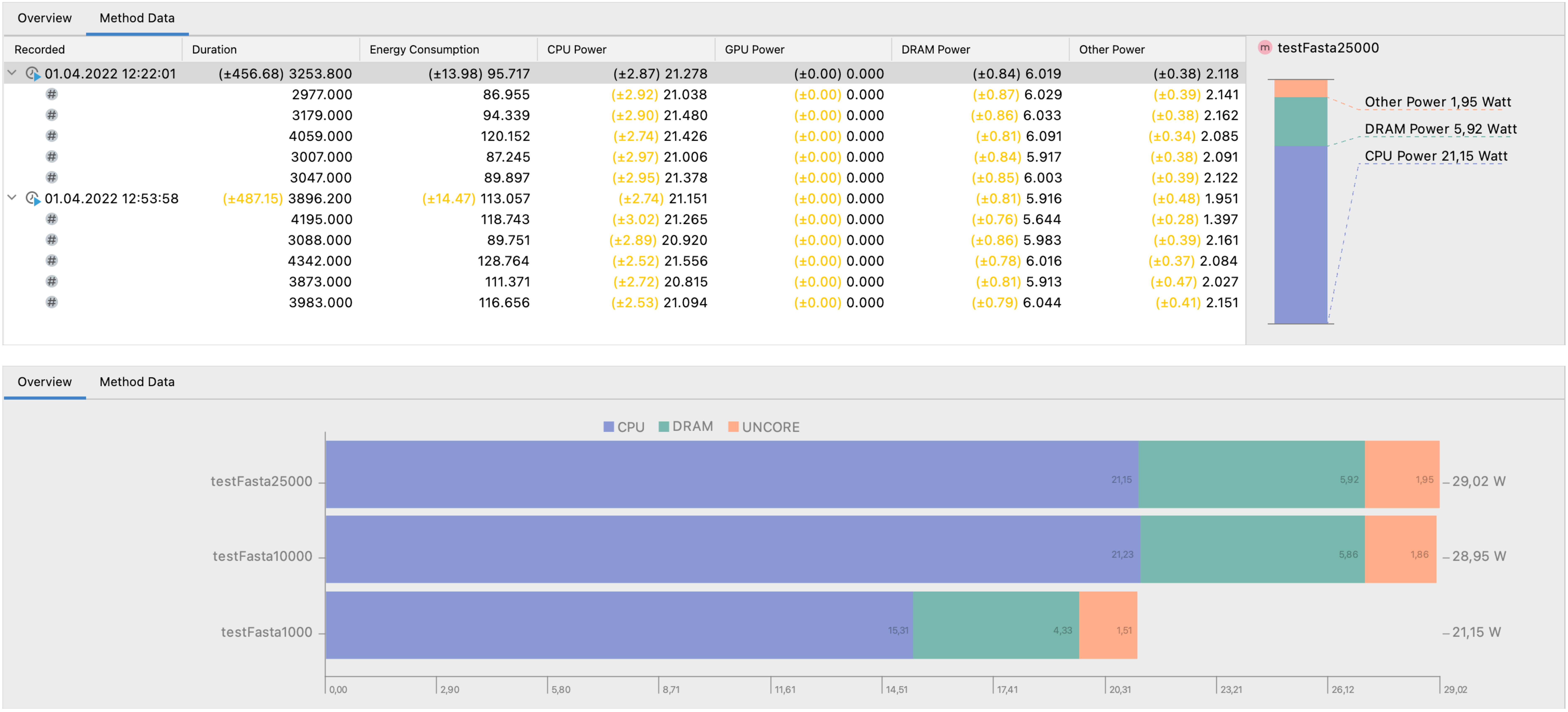}
    \caption{Visual comparison of the energy consumption of different unit test methods via the \emph{Summary Window}.}
    \label{fig:summary}
    \vspace{-1.6em}
\end{figure}

\paragraph{Summary Window} MANAi supports a detailed over\-view 
of the energy recordings, either represented as bar chart or 
using a contextual table view for the currently opened class (cf. Fig. \ref{fig:summary}). 
The \emph{Summary Window} offers the possibility to compare different 
unit tests within the same class. The data presented consists of the initially specified sampling rate 
the number of samples, the recorded power and energy data. 
Power and energy consumption data is reported for its individual domains 
as specified by Intel RAPL \cite{Liu2015DataOrientedCO,Pereira2021,5599016}.

\section{Conclusion}
In this paper, we presented a brief overview on 
MANAi an IntelliJ plugin for software energy profiling.
We believe that MANAi contributes to the ongoing challenges
around software sustainability by making energy characteristics 
of individual unit test methods in code explicit by providing visual feedback within 
the IDE. This way, our approach is beneficial for both education and development purposes. 
One of MANAi's limitations is a decrease in accuracy
if an instrumented test has a duration smaller than 
RAPL's update rate. However, related work on RAPL has already
provided possible solutions which are worthwhile being
investigated in depth \cite{10.1145/2425248.2425252}. An additional 
open challenge is the distortion of results due to RAPL's measurement 
overhead (cf. Desrochers et al.\cite{10.1145/2989081.2989088}). 
MANAi is hosted on an open-source GitHub 
repository\footnote{MANAi GitHub Repository -- https://github.com/aschuler84/manai}. 
A prepackaged binary of MANAi can be obtained from \cite{schuler_andreas_icts}. 
For future work, we seek to extend the plugin's support
for other programming languages like Kotlin, which 
due to its support for Java Bytecode and raising popularity amongst developers
makes it a promising candidate for integration.

\bibliographystyle{IEEEtran}
\balance
\bibliography{literature}

\end{document}